\documentclass[ALICE,manyauthors]{cernphprep}
\usepackage[comma,square,numbers,sort&compress]{natbib}
\usepackage{lineno}
\usepackage{xspace}
\usepackage{hyperref}
\usepackage{siunitx}
\usepackage[T1]{fontenc}
\DeclareSIUnit\clight{\text{\ensuremath{c}}}
\DeclareSIUnit\eVc{\eV/\clight} 
\DeclareSIUnit\keVc{\keV/\clight} 
\DeclareSIUnit\MeVc{\MeV/\clight} 
\DeclareSIUnit\MeVcc{\MeV/\clight\squared} 
\DeclareSIUnit\GeVc{\GeV/\clight} 
\DeclareSIUnit\GeVcc{\GeV/\clight\squared} 
\usepackage[Symbol]{upgreek}

\begin{document}
%

\newcommand{\pp}           {pp\xspace}

\newcommand{\roots}        {\ensuremath{\sqrt{s}}\xspace}
\newcommand{\pt}           {\ensuremath{p_{\rm T}}\xspace}
\newcommand{\px}           {\ensuremath{p_{\rm x}}\xspace}
\newcommand{\py}           {\ensuremath{p_{\rm y}}\xspace}
\newcommand{\etarange}[1]  {\mbox{$\left| \eta \right| < #1$}}
\newcommand{\yrange}[1]    {\mbox{$\left| y \right| < #1$}}
\newcommand{\dndy}         {\ensuremath{\mathrm{d}N_\mathrm{ch}/\mathrm{d}y}\xspace}
\newcommand{\dndeta}       {\ensuremath{\mathrm{d}N_\mathrm{ch}/\mathrm{d}\eta}\xspace}
\newcommand{\avdndeta}     {\ensuremath{\langle\dndeta\rangle}\xspace}
\newcommand{\dNdy}         {\ensuremath{\mathrm{d}N_\mathrm{ch}/\mathrm{d}y}\xspace}
\newcommand{\dEdx}         {\ensuremath{\textrm{d}E/\textrm{d}x}\xspace}
\newcommand{\kstar}        {\ensuremath{k^*}\xspace}
\newcommand{\mt}           {\ensuremath{m_{\rm{T}}}\xspace}
\newcommand{\St}           {\ensuremath{S_{\rm{T}}}\xspace}

\newcommand{\seven}        {$\sqrt{s}~=~7$~Te\kern-.1emV\xspace}
\newcommand{\thirteen}        {$\sqrt{s}~=~13$~Te\kern-.1emV\xspace}
\newcommand{\tev}          {Te\kern-.1emV\xspace}
\newcommand{\lumi}         {\ensuremath{\mathcal{L}}\xspace}

\newcommand{\ITS}          {\rm{ITS}\xspace}
\newcommand{\TOF}          {\rm{TOF}\xspace}
\newcommand{\ZDC}          {\rm{ZDC}\xspace}
\newcommand{\ZDCs}         {\rm{ZDCs}\xspace}
\newcommand{\ZNA}          {\rm{ZNA}\xspace}
\newcommand{\ZNC}          {\rm{ZNC}\xspace}
\newcommand{\SPD}          {\rm{SPD}\xspace}
\newcommand{\SDD}          {\rm{SDD}\xspace}
\newcommand{\SSD}          {\rm{SSD}\xspace}
\newcommand{\TPC}          {\rm{TPC}\xspace}
\newcommand{\TRD}          {\rm{TRD}\xspace}
\newcommand{\VZERO}        {\rm{V0}\xspace}
\newcommand{\VZEROA}       {\rm{V0A}\xspace}
\newcommand{\VZEROC}       {\rm{V0C}\xspace}
\newcommand{\Vdecay} 	   {\ensuremath{V^{0}}\xspace}

\newcommand{\ee}           {\ensuremath{\mathrm{e}^{+}\mathrm{e}^{-}}} 
\newcommand{\pip}          {\ensuremath{\pi^{+}}\xspace}
\newcommand{\pim}          {\ensuremath{\pi^{-}}\xspace}
\newcommand{\kap}          {\ensuremath{\mathrm{K}^{+}}\xspace}
\newcommand{\kam}          {\ensuremath{\mathrm{K}^{-}}\xspace}
\newcommand{\KK}           {\ensuremath{\mathrm{K}^{+}\mathrm{K}^{-}}\xspace}
\newcommand{\KKbar}           {\ensuremath{\mathrm{K}\overline{\mathrm{K}}}\xspace}
\newcommand{\phiPart}      {\ensuremath{\phi}\xspace}
\newcommand{\pbar}         {\ensuremath{\mathrm\overline{p}}\xspace}
\newcommand{\kzero}        {\ensuremath{{\mathrm K}^{0}_{\mathrm{S}}}\xspace}
\newcommand{\lmb}          {\ensuremath{\Lambda}\xspace}
\newcommand{\almb}         {\ensuremath{\overline{\Lambda}}\xspace}
\newcommand{\Om}           {\ensuremath{\Omega^-}\xspace}
\newcommand{\Mo}           {\ensuremath{\overline{\Omega}^+}\xspace}
\newcommand{\X}            {\ensuremath{\Xi^-}\xspace}
\newcommand{\Ix}           {\ensuremath{\overline{\Xi}^+}\xspace}
\newcommand{\Xis}          {\ensuremath{\Xi^{\pm}}\xspace}
\newcommand{\Oms}          {\ensuremath{\Omega^{\pm}}\xspace}

\newcommand{\Led}         {Lednick\'y--Lyuboshits\xspace}
\newcommand{\Ledn}         {Lednick\'y--Lyuboshits approach\xspace}
\newcommand{\Nphi}         {\ensuremath{\mathrm{N}}\mbox{--}\ensuremath{\phi}\xspace}
\newcommand{\pphi}         {\ensuremath{\mathrm{p}}\mbox{--}\ensuremath{\phi}\xspace}
\newcommand{\apphi}        {\ensuremath{\overline{\mathrm{p}}}\mbox{--}\ensuremath{\phi}\xspace}
\newcommand{\pphiComb}     {\ensuremath{\mathrm{p}}\mbox{--}\ensuremath{\phi} \ensuremath{\oplus} \ensuremath{\overline{\mathrm{p}}}\mbox{--}\ensuremath{\phi}\xspace}

\newcommand{\pKK}         {\ensuremath{\mathrm{p}}\mbox{--}(\ensuremath{\mathrm{K}^{+}\mathrm{K}^{-}})\xspace}

\newcommand{\pP}           {\ensuremath{\mathrm{p}\mbox{--}\mathrm{p}}\xspace}
\newcommand{\ApAP}         {\ensuremath{\pbar\mbox{--}\pbar}\xspace}
\newcommand{\pPComb}       {\ensuremath{\mathrm{p}\mbox{--}\mathrm{p} \oplus \pbar\mbox{--}\pbar}\xspace}

\newcommand{\fReal} {\ensuremath{\Re(f_0) =\nolinebreak 0.85 \pm\nolinebreak 0.34\,(\mathrm{stat.}) \pm\nolinebreak 0.14\,(\mathrm{syst.})~\si{fm}}\xspace}
\newcommand{\fImag} {\ensuremath{\Im(f_0) =\nolinebreak 0.16 \pm\nolinebreak 0.10\,(\mathrm{stat.}) \pm\nolinebreak 0.09\,(\mathrm{syst.})~\si{fm}}\xspace}
\newcommand{\dZero} {\ensuremath{d_0 =\nolinebreak 7.85 \pm\nolinebreak 1.54\,(\mathrm{stat.})  \pm\nolinebreak 0.26\,(\mathrm{syst.})~\si{fm}}\xspace}

\newcommand{\NsigmaNoCorr} {\ensuremath{5.7 \pm\nolinebreak 0.8\,(\mathrm{stat.})  \pm\nolinebreak 0.5\,(\mathrm{syst.})~\sigma}\xspace }

\newcommand{\NsigmaNoCorrRange} {\ensuremath{4.7 - 6.6~\sigma}\xspace}

\newcommand{\VeffGauss} {\ensuremath{V_{\mathrm{eff}}=\nolinebreak 2.5\pm\nolinebreak 0.9 \,(\mathrm{stat.})\pm\nolinebreak 1.4\,(\mathrm{syst.})~\si{\MeV}}\xspace}
\newcommand{\muGauss}   {\ensuremath{\upmu=\nolinebreak0.14\pm\nolinebreak 0.06 \,(\mathrm{stat.})\pm\nolinebreak 0.09\,(\mathrm{syst.})~\si{fm^{-2}}}\xspace}

\newcommand{\alphaYuk}  {\ensuremath{\upalpha=\nolinebreak65.9 \pm\nolinebreak 38.0\,(\mathrm{stat.})\pm\nolinebreak 17.5\,(\mathrm{syst.})~\si{\MeV}}\xspace}
\newcommand{\AYuk}  {\ensuremath{A=\nolinebreak0.021\pm\nolinebreak 0.009\,(\mathrm{stat.})\pm\nolinebreak 0.006\,(\mathrm{syst.})}\xspace}
\newcommand{\gYuk}  {\ensuremath{g_{\Nphi} =\nolinebreak 0.14\pm\nolinebreak 0.03\,(\mathrm{stat.})\pm\nolinebreak 0.02\,(\mathrm{syst.})}\xspace}

\newcommand{\norm} {\ensuremath{\mathit{M} = 0.96}\xspace}

\newcommand{\realScatSigma}{\ensuremath{2.3\upsigma}\xspace}

\begin{titlepage}
\PHyear{2021}       
\PHnumber{081}      
\PHdate{10 May}  

\title{Experimental evidence for an attractive \pphi interaction}
\ShortTitle{Experimental evidence for an attractive \pphi interaction}   

\Collaboration{ALICE Collaboration\thanks{See Appendix~\ref{app:collab} for the list of collaboration members}}
\ShortAuthor{ALICE Collaboration} 

\begin{abstract}
This Letter presents the first experimental evidence of the attractive strong interaction between a proton and a \phiPart meson. The result is obtained from two-particle correlations of combined \pphiComb pairs measured in high-multiplicity \pp collisions at \thirteen by the ALICE collaboration.
The spin-averaged scattering length and effective range of the \pphi interaction are extracted from the fully corrected correlation function employing the \Ledn.
In particular, the imaginary part of the scattering length vanishes within uncertainties, indicating that inelastic processes do not play a prominent role for the \pphi interaction.
These data demonstrate that the interaction is dominated by elastic \pphi scattering.
Furthermore, an analysis employing phenomenological Gaussian- and Yukawa-type potentials is conducted. Under the assumption of the latter, the \Nphi coupling constant is found to be \gYuk.
This work provides valuable experimental input to accomplish a self-consistent description of the \Nphi interaction,  which is particularly relevant for the more fundamental studies on partial restoration of chiral symmetry in nuclear medium.

\end{abstract}
\end{titlepage}

\setcounter{page}{2} 


Quantum Chromodynamics (QCD) is the theory of the strong interaction, where the degrees of freedom are represented by colored quarks and gluons. 
The observable degrees of freedom at low energies, however, are composite hadrons, and their mass is dynamically generated by the interplay between spontaneous and explicit chiral symmetry breaking. The spontaneous breaking arises from the non-zero vacuum expectation value $\langle 0 | \mathrm{\overline{q}}\mathrm{q} | 0\rangle$ of the ground state of the QCD Lagrangian, the quark condensate $\mathrm{\overline{q}}$q. The consideration of the quark mass term leads to the explicit breaking of chiral symmetry.
A link between the hadron properties and $\langle 0 | \mathrm{\overline{q}}\mathrm{q} | 0 \rangle$ is provided by the QCD sum rule method~\cite{SumRuleHatsuda,Leupold:2009kz,Zschocke:2002mn,Brown:1991kk}, with the hadrons being interpreted as excitations of the $\mathrm{\overline{q}}$q ground state. 
Chiral symmetry is expected to be partially restored in a dense and/or hot strongly interacting medium~\cite{BROCKMANN199640}. 
Consequently, the modification of the $\mathrm{\overline{q}}$q condensate in the medium is reflected in the spectral shape of its hadronic excitations. This results in a potentially measurable mass shift and/or width broadening of the hadrons. 
From the experimental point of view, the light vector mesons $\uprho$, $\upomega$ and \phiPart represent the most suitable hadronic probes due to the short lifetime~\cite{Rapp:1999ej, Metag:2017ixh}. The detection through the dileptonic decay channels (\ee and $\upmu^+\upmu^-$) allows to infer on the meson properties at the time of their decay~\cite{CERES, NA60, Agakishiev:2011vf, PHENIX, STAR}. 

In this context, the \phiPart meson represents an experimentally challenging probe due to the narrow decay width in vacuum, which renders its spectral shape distinguishable from that of the $\uprho$ and $\upomega$ mesons. The KEK-PS E325 collaboration measured the \phiPart spectral function in p--C and p--Cu reactions and reported a slight mass shift and moderate increase of the width, which was in a model-dependent manner interpreted as an in-medium modification~\cite{Muto:2005za}. 
The collected data were, however, limited due to the small branching ratio (BR $\approx \num{3e-4}$~\cite{PDG}) of the $\phiPart \rightarrow \ee$ decay.
Larger data samples were collected by the Spring8-LEPS collaboration in order to determine the \phiPart production and absorption cross section in photon-induced reactions by exploiting the decay to \KK (BR $\approx 50 \%$~\cite{PDG})~\cite{Ishikawa:2004id}. 
The \Nphi cross section in nuclear matter was found to be much larger than the vacuum cross section, corresponding to an in-medium \phiPart width of about \SI{110}{\MeVc^2}. A similar conclusion was reached by other experiments from transparency ratio measurements in photon-~\cite{CLAS_absorp} and proton-induced reactions~\cite{ANKE_absorp}.

The interpretation of these results is far from trivial due to the lack of model-independent information on the \Nphi interaction and its complex dynamics. Because of the hidden strangeness content ($\mathrm{s}\overline{\mathrm{s}}$) of the \phiPart, the direct coupling to u and d quarks in the nucleon is expected to be suppressed due to the Okubo-Zweig-Iizuka (OZI) rule~\cite{OZI_O, OZI_Z, OZI_I}. 
Indications for a possible OZI rule violation were reported by the HADES collaboration~\cite{Agakishiev:2011vf}.
In addition, several scenarios allow for a direct interaction of the \phiPart meson with the nucleon, including a coupling to its strangeness content~\cite{Ellis:1988jwa}. 
An alternative is the QCD Van der Waals interaction~\cite{QCDvdW}, mediated by multi-gluon exchange, which has been identified as relevant for the ($\mathrm{c}\overline{\mathrm{c}}$)--N system~\cite{QCDvdWCharm}. Using a phenomenological Yukawa-type potential with adjusted parameters, the QCD Van der Waals force is able to form a \Nphi bound state~\cite{Gao:2000az}. However, the direct \Nphi coupling is typically not considered in theoretical descriptions of the interaction in vacuum and in medium. 
OZI-allowed \Nphi processes can also proceed via the coupling to particle pairs with same quantum numbers, such as $\mathrm{K\mbox{--}\Uplambda, K\mbox{--}\Upsigma, K^*\mbox{--}\Uplambda, K^*\mbox{--}\Upsigma}$~\cite{Cabrera:2017agk, PhysRevC.95.015201, Klingl:1997tm}. 
More importantly, the properties of the \phiPart are defined by the $\phiPart-\KKbar$ coupling~\cite{OZI_O, OZI_Z, OZI_I}. Accordingly, \KKbar loops and tadpole diagrams contribute to the self-energy of the \phiPart~\cite{Gubler:2016itj, PhysRevC.95.015201, Klingl:1997tm}, linking the \phiPart properties to the complex dynamics of the K\mbox{--}N and $\overline{\mathrm{K}}$\mbox{--}N systems~\cite{PhysRevLett.124.132501}. This was explored by HADES in $\pi$\mbox{--}nucleus reactions at beam energies of \SI{1.65}{GeV}, demonstrating a constant $\phiPart/\kam$ ratio in different collision systems~\cite{HADES_absorp}. Since a strong \kam absorption was observed also a strong \phiPart absorption is expected. 
The overall \Nphi interaction is defined by the interplay of all aforementioned processes. However, the present theoretical and experimental situation is far from being resolved~\cite{TolosRev}. 
Indeed, a consistent description of the available photon- and proton-induced data~\cite{Ishikawa:2004id, CLAS_absorp, ANKE_absorp} is still out of reach~\cite{Cabrera:2017agk}.
This demonstrates the need for a direct measurement of the two-body \Nphi interaction in vacuum, in order to constrain theoretical models and correctly interpret the data from nuclear collisions.

Recently, detailed measurements of the final-state interaction were conducted employing two-particle correlations in ultra-relativistic proton–proton (\pp) collisions ~\cite{FemtoRun1, FemtoSource, FemtoLambdaLambda, FemtoKaon, FemtopXi, FemtopSigma, FemtoNature, FemtoRev}. In this Letter, these studies are extended to protons and \phiPart mesons.

The observable is the two-particle correlation function $C(\kstar)$, defined as $C(\kstar)=\mathcal{N}\times\frac{N_{\rm{same}}(\kstar)}{N_{\rm{mixed}}(\kstar)}$ \cite{Lisa:2005dd} 
where $N_{\rm{same}}$ and $N_{\rm{mixed}}$ are the distributions of the relative momentum $\kstar = \frac{1}{2}\times|\mathbf{p}^*_1-\mathbf{p}^*_2|$ between both particles in the pair rest frame (denoted by the $^*$) and a normalization constant $\mathcal{N}$, evaluated from same and mixed events, respectively. The Koonin-Pratt equation $C(\kstar)=\int \mathrm{d}^3 r^{*} ~ S(r^{*}) \mid \Psi(\vec{r^{*}}, \vec{\kstar}) \mid ^2$~\cite{Lisa:2005dd} relates the observable to the source function $S(r^*)$ and the two-particle wave function $\Psi(\vec{r^{*}}, \vec{\kstar})$ incorporating the final-state interaction, where $r^{*}$ is the relative distance between the two particles.

The results reported in this Letter are based on the analysis of a data sample of \pp collisions at \thirteen recorded by ALICE~\cite{ALICE, ALICEperf} during the LHC Run 2 (2015--2018). A high-multiplicity trigger relying on the measured signal amplitudes in the \VZERO detectors~\cite{VZERO} is employed to select collisions with on average 30 produced charged particles in the pseudorapidity interval \etarange{0.5}~\cite{FemtoNature}. The resulting data sample represents the upper 0.17\% of the charged-particle distribution of all inelastic collisions with at least one charged particle in the range $|\eta| < 1$ (referred to as INEL $>$ 0)~\cite{FemtopSigma}. The event selection follows~\cite{FemtopSigma, FemtoRun1}. The reconstructed primary vertex is required to be located within $\pm$\SI{10}{cm} of the nominal interaction point along the beam direction to assure a uniform detector coverage.
Meson-baryon correlation measurements in \pp collisions are contaminated by a so-called minijet background, induced by jet-like structures associated with hard parton-parton scatterings~\cite{minijets, FemtoKaon}, which influence the event shape~\cite{Sphericity1, Sphericity2}. Therefore, a selection on the transverse sphericity \St~\cite{Sphericity1, Sphericity2, FemtoKaon}, defined as in Ref.~\cite{Sphericity1}, is utilized to reduce the minijet contribution.
A total of \num{5e8} events with $0.7 < \St < 1.0$~\cite{FemtoKaon} are analyzed.
The subsystems employed in the analysis are the Inner Tracking System (ITS)~\cite{ALICE}, the Time Projection Chamber (TPC)~\cite{TPC} and the Time-Of-Flight (\TOF) detector~\cite{TOF}, covering the full azimuthal angle and the pseudorapidity interval \etarange{0.9}. The detectors are immersed in a uniform magnetic field of \SI{0.5}{\tesla} along the beam direction.

The proton candidates are selected following the methods used in Ref.~\cite{FemtoRun1}. The particle identification (PID) is conducted employing the TPC and TOF detectors by determining the deviation $n_\sigma$ between the signal hypothesis for a proton and the measurement, normalized by the detector resolution $\sigma$.
Contributions from secondary particles stemming from weak decays or the interaction of primary particles with the detector material are extracted using Monte Carlo (MC) template fits to the measured distribution of the Distance of Closest Approach (DCA) of the track to the primary vertex~\cite{FemtoRun1}. This results in a proton purity of 99\%, with a primary fraction of 82\%~\cite{FemtoSource}. 
The \phiPart meson is reconstructed from its hadronic decay to charged kaons $\phiPart \rightarrow \KK$ as in Refs.~\cite{Abelev_2012, Acharya_2020, Acharya:2019bli}.
Charged kaons are identified with a transverse momentum of $\pt > \SI{0.15}{\GeVc}$ and \etarange{0.8}. For momenta $p < \SI{0.4}{\GeVc}$ the PID selection provided by the TPC is used, while for larger momenta the information of TPC and TOF is combined. A selection on the DCA of the track to the primary vertex in both the beam direction  ($|\mathrm{DCA}_{z}|<\SI{0.8}{cm}$) and transverse plane ($|\mathrm{DCA}_{xy}|<\SI{0.4}{cm}$) is employed to increase the fraction of kaons from \phiPart decays. The purity of the kaon sample is $>90\%$ for $\pt < \SI{1.25}{\GeVc}$. The \KK invariant mass is calculated combining two oppositely charged kaons assuming their nominal masses~\cite{PDG}. The \phiPart candidates are selected within a window of $\pm \SI{8}{\MeVcc}$ around the nominal \phiPart mass, resulting in a total number of \num{5.8e6} with a purity of $66\%$.

\begin{figure}
\centering
\includegraphics[width=0.75\textwidth]{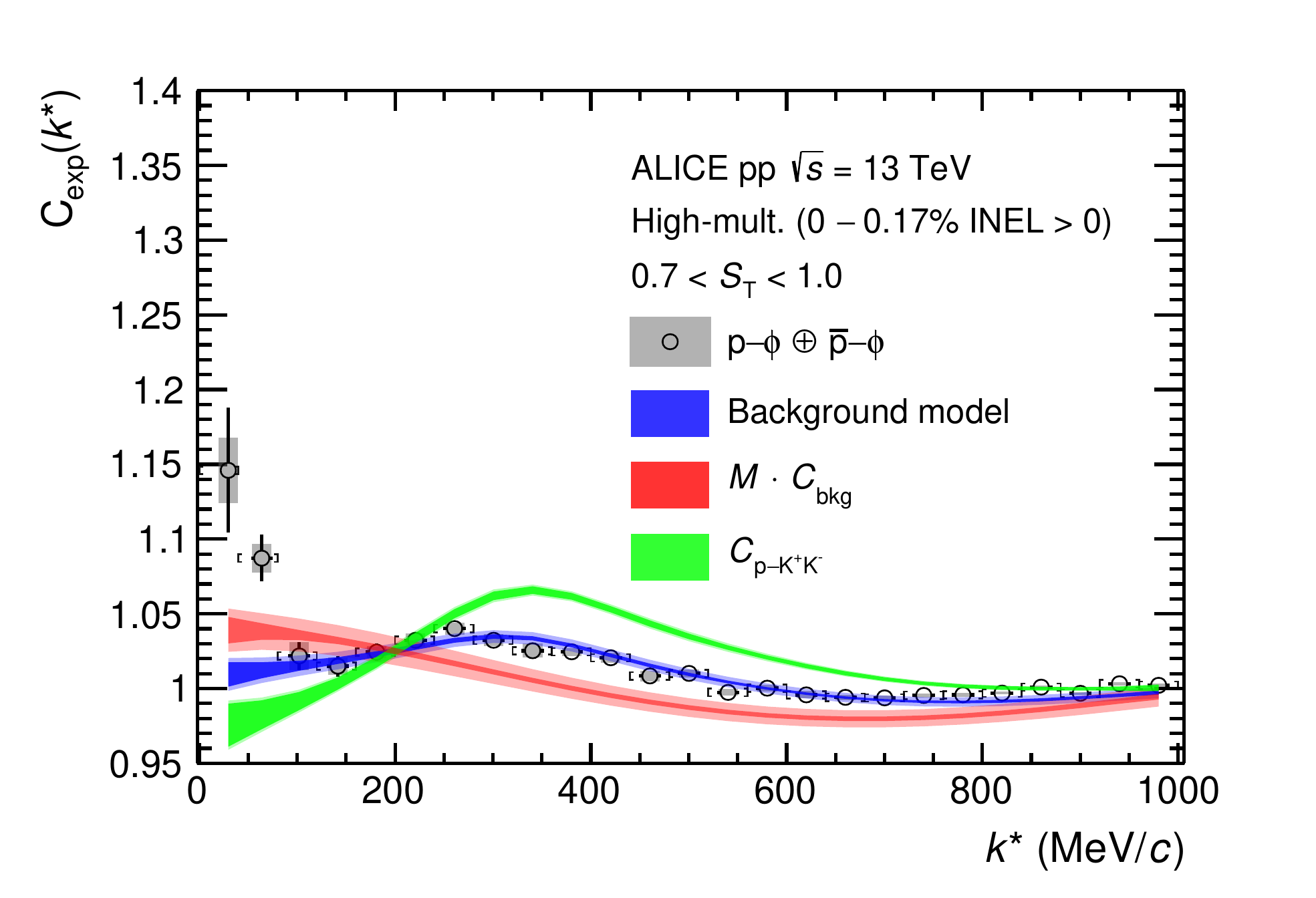}
\caption{The experimental \pphi correlation function $C_{\mathrm{exp}}(\kstar)$ and various contributions as described in Eq.~\ref{eq:tot}. Statistical (bars) and systematic uncertainties (boxes) are shown separately. The width of the dark (light) shaded bands depicts the statistical (total) uncertainty.}
\label{fig:contributions}
\end{figure}

A total of $\num{4.17e4}$ \pphi and $\num{3.61e4}$ \apphi pairs with $\kstar<\SI{200}{\MeVc}$ contribute to the $N_{\rm{same}}$ distribution of the respective correlation function. As they are compatible within uncertainties, both correlation functions are combined. In the following, \pphi refers to \pphiComb. The correlation function is normalized within $\kstar\in[800,1000]~\si{\MeVc}$, where all the contributions to the correlation function are expected to be flat. The data are unfolded for the finite momentum resolution of the detector~\cite{ROOUNFOLD}, which modifies the correlation function by 0.7\% at low \kstar. 
The measured \pphi correlation function is shown in Fig.~\ref{fig:contributions}.
The \kstar value of the data points is determined by the average  $\langle$\kstar$\rangle$ of the $N_{\mathrm{same}}$ distribution in the corresponding interval. 
In the region $150 < \kstar < \SI{600}{\MeVc}$ a rise in the correlation function attributed to residual minijet background is visible. Therefore, any conclusion on the genuine \pphi interaction demands a treatment of the contributions due to residual correlations from the underlying event topology (minijets and energy/momentum conservation), weak decays feeding to the particles of interest (feed-down) and misidentifications~\cite{FemtoRun1}. 

The experimental \pphi correlation function $C_{\mathrm{exp}}(\kstar)$ is decomposed as
\begin{equation}
C_{\mathrm{exp}}(\kstar) = \mathit{M} \times C_{\mathrm{bkg}}(\kstar)  \times \left[\lambda_{\pphi} \times C_{\pphi}(\kstar) + \lambda_{\mathrm{flat}} \times C_{\mathrm{flat}}(\kstar) \right]+ \lambda_{\pKK} \times C_{\pKK}(\kstar),
\label{eq:tot}
\end{equation}
where $\mathit{M}$ is a normalization constant, $C_{\mathrm{bkg}}(\kstar)$ is the non femtoscopic background, $C_{\pphi}(\kstar)$ the genuine \pphi correlation function, $C_{\mathrm{flat}}(\kstar) \approx 1$ considers the effect from weak decays and single particle misidentification and $C_{\pKK}(\kstar)$ is the contribution from combinatorial \KK background. Correlations are summed for independent pairs and multiplied when considering different effects acting on the same pair. The parameters $\lambda_i$, summarized in Table~\ref{tableLambdaParam}, are obtained in a data-driven way from single-particle properties~\cite{FemtoRun1}. As the purity of the \phiPart candidates depends on \pt, it is evaluated for those entering the correlation function, and found to be 57\%.

The combinatorial \KK background, referred to as \pKK, significantly contributes to the measured correlation function. Its shape is extracted from the sidebands of the \KK invariant mass selection and mainly driven by p\mbox{--}\kap and p\mbox{--}\kam interactions. The sideband intervals are chosen as $0.995-1.011~\si{\giga\eV\per\clight^2}$ and $1.028-1.044~\si{\giga\eV\per\clight^2}$ to avoid threshold effects and have comparable kinematic properties as the \phiPart candidates. The resulting correlation function is parametrized with a double Gaussian and a quadratic polynomial. Finally, a residual \phiPart amount of $8.6\%$ in the sidebands is considered, which arises from the tail of the \phiPart resonance extending into the sideband intervals. This results in a $7\%$ contribution to the experimental $C_{\mathrm{\pKK}}(\kstar)$, which is absorbed by a renormalization of the $\lambda$ parameters. Since the \pKK contribution is obtained from data, the corresponding residual minijet background and energy-momentum conservation effects are accounted for. The resulting $C_{\mathrm{\pKK}}(\kstar)$ is depicted by the green band in Fig.~\ref{fig:contributions}.

The non femtoscopic background $C_{\mathrm{bkg}}(\kstar)$ in Eq.~\ref{eq:tot} is multiplied to the genuine \pphi and flat contributions. 
$C_{\mathrm{bkg}}(\kstar) = C_{\mathrm{baseline}}(\kstar) + C_{\mathrm{minijet}}(\kstar)$ is dominated by residual minijet contributions $C_{\mathrm{minijet}}(\kstar)$ obtained from \textsc{Pythia 8}~\cite{PYTHIA} generated events, which yield a consistent description of such background~\cite{minijets1, minijets2}. However, further correlations at large \kstar due to energy-momentum conservation effects are not properly reproduced by \textsc{Pythia 8}~\cite{ALICE:2016jjg}. Therefore, an additional $C_\mathrm{baseline}(\kstar)$ function, described by a second order polynomial, is considered. 

\begin{table}   
\centering
\caption[lambda parameters]{Weight parameters of the individual components of the \pphi correlation function.}
\begin{tabular}{l | c}
\textbf{Pair} & \textbf{Weight} $\mathbf{\lambda}$ \textbf{(\%)} \\
\hline
\pphi & 46.3 \\
\pKK & 43.3 \\
Flat &  10.4 \\
\end{tabular} 
\label{tableLambdaParam}
\end{table}

The data is then fitted with a background model (Eq.~\ref{eq:tot} with $C_{\pphi}(\kstar) = 1$) within $\kstar\in[200, 800]~\si{\MeVc}$, which accounts for all contributions besides the genuine \pphi interaction. The $\lambda$ parameters and the shape of the minijet and \pKK background are fixed in the fit. Only the normalization constant and baseline parameters are free. The resulting $C_{\mathrm{bkg}}(\kstar)$ is shown by the red band in Fig.~\ref{fig:contributions} and the mean value of the normalization constant is \norm. Accordingly, the total background is obtained as the blue band in Fig.~\ref{fig:contributions}, which accurately reproduces the enhancement between $\SI{200}{\MeVc}$ and $\SI{1000}{\MeVc}$. 

\begin{figure}
\centering
\includegraphics[width=0.75\textwidth]{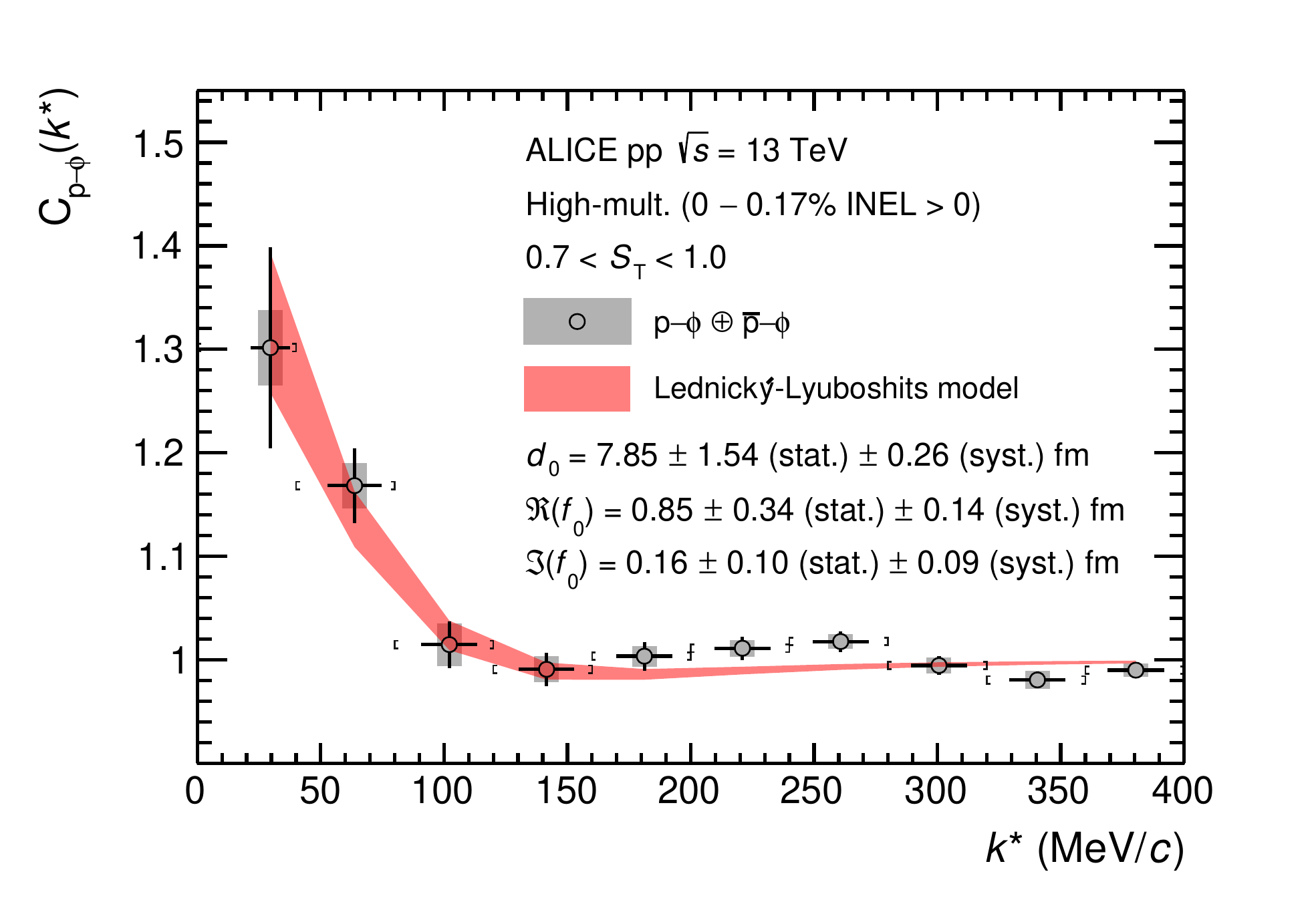}
\caption{The genuine \pphi correlation function $C_{\pphi}(\kstar)$ with statistical (bars) and systematic uncertainties (boxes). The red band depicts the results from the fit employing the \Ledn~\cite{Lednicky:1981su}. The width corresponds to one standard deviation of the uncertainty of the fit.}
\label{fig:Lednickyfit}
\end{figure}

The genuine \pphi contribution $C_{\pphi}(\kstar)$, is extracted from Eq.~\ref{eq:tot} by subtracting the obtained background contributions from the experimental correlation and is shown in Fig.~\ref{fig:Lednickyfit}. For $\kstar > \SI{200}{\MeVc}$, it is flat and consistent with unity within uncertainties. At low \kstar a pronounced enhancement with a significance of \NsigmaNoCorrRange with respect to the unity becomes apparent, which evidences the attractive nature of the \pphi interaction. 
The systematic uncertainties of the data and the background description are assessed by varying simultaneously the selection criteria of protons and kaons as well as the lower limit of the sphericity. These variations are chosen such that any combination leads to a maximum change of $\pm20\%$ of $N_{\mathrm{same}}$ within $\kstar < \SI{200}{\MeVc}$ in order to retain the statistical significance. Systematic uncertainties associated with the background description are evaluated by varying the fit ranges and the order of the polynomial assumed for $C_\mathrm{baseline}(\kstar)$. Uncertainties related to the unfolding are accounted for according to Ref.~\cite{FemtoNature}. This results in a relative systematic uncertainty at low \kstar of $2.8\%$.

In correlation measurements, the detected pairs are emitted in the final state of the scattering processes. The correlation function of the sample is then sensitive to elastic and inelastic channels produced in the collision~\cite{HaidenbauerCC}.
Inelastic channels opening below threshold act as an effective increase of the correlation function. The relevant channels for the \pphi system, $\Uplambda\mbox{--}\mathrm{K}$ and $\Upsigma\mbox{--}\mathrm{K}$ are located substantially below threshold.
Channels appearing above threshold lead to a cusp structure in $C(\kstar)$ in the vicinity of the threshold. Because of the large uncertainties and the broad bin width, no such structures are observed at the opening of the $\Uplambda\mbox{--}\mathrm{K}^{*}$ ($\kstar = \SI{221.6}{\MeVc}$) and $\Upsigma\mbox{--}\mathrm{K}^{*}$ ($\kstar = \SI{357.4}{\MeVc}$) thresholds. 

In order to interpret the measured genuine \pphi correlation one has to consider that the \pphi interaction features one isospin and two spin configurations. Since the latter cannot be disentangled, spin-averaged results are presented.
The strong \pphi interaction is modeled employing the \Ledn~\cite{Lednicky:1981su}. Coupled channel effects are incorporated via an imaginary contribution to the scattering length. 
For large values of $d_0$, the term $\propto d_0 / r_0$ that corrects the asymptotic wave function for small sources has an impact on the modeled correlation function~\cite{FemtoLambdaLambda}.
Additionally, in line with studies of charmonium states~\cite{QCDvdWCharm, Yokota:2013sfa}, phenomenological potentials are employed to model the \pphi interaction~\cite{Gao:2000az}, including Yukawa-type, $V_{\mathrm{Yukawa}}(r)=-A\times r^{-1}\times  e^{-\upalpha\times r}$, and Gaussian-type $V_{\mathrm{Gaussian}}(r)=-V_{\mathrm{eff}}\times e^{-\upmu\times r^2}$ potentials. The correlation functions based on these potentials are obtained with \textit{The Correlation Analysis Tool using the Schr\"odinger Equation} (CATS)~\cite{CATS}.

The particle-emitting source is extracted from studies of \pP and p\mbox{--}$\Uplambda$ pairs~\cite{FemtoSource}, which demonstrated that by accounting for the effect of strong resonances feeding to the particle pair of interest, a common source for both pairs is found. The primordial source depends on the transverse mass \mt of the particle pair and is obtained by evaluating the core radius at the $\langle\mt\rangle = \SI{1.66}{\GeVcc}$ of the \pphi pairs. The strong decays feeding to protons are explicitly considered~\cite{FemtoSource}, while for the \phiPart a 100\% primordial fraction is assumed~\cite{PDG}. The resulting source function is parametrized by a Gaussian profile with $r_{\mathrm{eff}}=(1.08\pm0.05)~\si{fm}$.

The interaction parameters are extracted by fitting the genuine \pphi correlation function $C_{\pphi}(\kstar)$ with the respective model within $\kstar < \SI{200}{\MeVc}$.
The systematic uncertainties of the procedure are assessed by varying the upper limit of the fit range by $\pm\SI{30}{\MeVc}$ and the source radius within its uncertainties.

 The real and imaginary parts of the scattering length obtained from the \Led fit are \fReal and \fImag. The resulting effective range is \dZero. $\Re(f_0)$ deviates by \realScatSigma from zero, indicating the attractiveness of the \pphi interaction in the approximate vacuum of \pp collisions.
Notably, $\Im(f_0)$ vanishes within uncertainties, indicating that inelastic processes do not play a prominent role in the interaction. Instead, the elastic \pphi interaction appears to be dominant in vacuum. 
The scattering length is larger than values found in literature: a recent analysis of data recorded with the CLAS experiment reports $|f_0| = (0.063 \pm 0.010)$~fm~\cite{Strakovsky:2020uqs}; a value of around $f_0 = \SI{0.15}{fm}$ is consistent with LEPS measurements of the \phiPart cross section~\cite{PhysRevC.76.048202, LEPS}; studies of an effective Lagrangian combining chiral SU(3) dynamics with vector meson dominance obtain $f_0 = (-0.01 + i\,0.08)$~fm~\cite{Klingl:1997kf}; and a QCD sum rule analysis finds $f_0 = (-0.15 \pm 0.02)$~fm~\cite{10.1143/PTP.98.631}. The obtained scattering lengths are rather model dependent since the data refer to the properties of the \phiPart meson inside a nucleus and not to a two-body system as in this work. This underlines the importance of direct measurements of the two-body \Nphi interaction to provide constraints for theoretical models.

Finally, the data are employed to constrain the parameters of phenomenological Gaussian- and Yukawa-type potentials. As the imaginary contribution of the scattering length is consistent with zero, only real values are used for the parameters. 
The fits yield a comparable degree of consistency as the fit with the \Ledn. 
The resulting values for the Gaussian-type potential are \VeffGauss and \muGauss, indicating a much shallower strong interaction potential than lattice QCD results for the N--$\mathrm{J}/\psi$ strong interaction~\cite{jpsi}.
For the Yukawa-type potential the fit yields \AYuk and \alphaYuk. Predictions of possible \Nphi bound states employing the same kind of potential with modified parameters  ($\upalpha = \SI{600}{MeV}$ and $A = 1.25$)~\cite{Gao:2000az} are therefore incompatible with this measurement. The \Nphi coupling constant under the assumption of a Yukawa-type potential ($g_{\Nphi} = \sqrt{A}$) is \gYuk.

In conclusion, this Letter presents the first correlation-based measurement of the \pphi interaction. The correlation function reflects the pattern of an attractive interaction. The scattering parameters, extracted with the \Ledn, are \fReal,\linebreak \fImag, and \dZero. Remarkably, the imaginary contribution to the scattering length vanishes, indicating that inelastic processes do not play a prominent role. 
Instead, the interaction is dominated by the elastic \pphi dynamics. Under the assumption of a Yukawa-type potential for the \Nphi interaction, the value of the coupling constant is extracted as \gYuk. \\
These results seem to contradict the interpretation of the \phiPart production off nuclear targets in terms of large absorption cross-sections but more data are needed to extract the precise value of the imaginary part of the scattering length and the here reported results should serve as an input for more advanced modellings in medium.
The upcoming Run 3 and Run 4 data taking at the LHC will allow to significantly improve the precision of the extracted interaction parameters.
This measurement demonstrates for the first time that the \pphi interaction in vacuum is attractive and dominated by elastic scattering. It provides valuable experimental input on the \Nphi interaction, which is fundamental to reach a self-consistent description of the interaction as required for the correct interpretation of data from nuclear collisions. 


\newenvironment{acknowledgement}{\relax}{\relax}
\begin{acknowledgement}
\section*{Acknowledgements}
The ALICE Collaboration is grateful to P.~Gubler and N.~Kaiser for valuable discussions.

The ALICE Collaboration would like to thank all its engineers and technicians for their invaluable contributions to the construction of the experiment and the CERN accelerator teams for the outstanding performance of the LHC complex.
The ALICE Collaboration gratefully acknowledges the resources and support provided by all Grid centres and the Worldwide LHC Computing Grid (WLCG) collaboration.
The ALICE Collaboration acknowledges the following funding agencies for their support in building and running the ALICE detector:
A. I. Alikhanyan National Science Laboratory (Yerevan Physics Institute) Foundation (ANSL), State Committee of Science and World Federation of Scientists (WFS), Armenia;
Austrian Academy of Sciences, Austrian Science Fund (FWF): [M 2467-N36] and Nationalstiftung f\"{u}r Forschung, Technologie und Entwicklung, Austria;
Ministry of Communications and High Technologies, National Nuclear Research Center, Azerbaijan;
Conselho Nacional de Desenvolvimento Cient\'{\i}fico e Tecnol\'{o}gico (CNPq), Financiadora de Estudos e Projetos (Finep), Funda\c{c}\~{a}o de Amparo \`{a} Pesquisa do Estado de S\~{a}o Paulo (FAPESP) and Universidade Federal do Rio Grande do Sul (UFRGS), Brazil;
Ministry of Education of China (MOEC) , Ministry of Science \& Technology of China (MSTC) and National Natural Science Foundation of China (NSFC), China;
Ministry of Science and Education and Croatian Science Foundation, Croatia;
Centro de Aplicaciones Tecnol\'{o}gicas y Desarrollo Nuclear (CEADEN), Cubaenerg\'{\i}a, Cuba;
Ministry of Education, Youth and Sports of the Czech Republic, Czech Republic;
The Danish Council for Independent Research | Natural Sciences, the VILLUM FONDEN and Danish National Research Foundation (DNRF), Denmark;
Helsinki Institute of Physics (HIP), Finland;
Commissariat \`{a} l'Energie Atomique (CEA) and Institut National de Physique Nucl\'{e}aire et de Physique des Particules (IN2P3) and Centre National de la Recherche Scientifique (CNRS), France;
Bundesministerium f\"{u}r Bildung und Forschung (BMBF) and GSI Helmholtzzentrum f\"{u}r Schwerionenforschung GmbH, Germany;
General Secretariat for Research and Technology, Ministry of Education, Research and Religions, Greece;
National Research, Development and Innovation Office, Hungary;
Department of Atomic Energy Government of India (DAE), Department of Science and Technology, Government of India (DST), University Grants Commission, Government of India (UGC) and Council of Scientific and Industrial Research (CSIR), India;
Indonesian Institute of Science, Indonesia;
Istituto Nazionale di Fisica Nucleare (INFN), Italy;
Institute for Innovative Science and Technology , Nagasaki Institute of Applied Science (IIST), Japanese Ministry of Education, Culture, Sports, Science and Technology (MEXT) and Japan Society for the Promotion of Science (JSPS) KAKENHI, Japan;
Consejo Nacional de Ciencia (CONACYT) y Tecnolog\'{i}a, through Fondo de Cooperaci\'{o}n Internacional en Ciencia y Tecnolog\'{i}a (FONCICYT) and Direcci\'{o}n General de Asuntos del Personal Academico (DGAPA), Mexico;
Nederlandse Organisatie voor Wetenschappelijk Onderzoek (NWO), Netherlands;
The Research Council of Norway, Norway;
Commission on Science and Technology for Sustainable Development in the South (COMSATS), Pakistan;
Pontificia Universidad Cat\'{o}lica del Per\'{u}, Peru;
Ministry of Education and Science, National Science Centre and WUT ID-UB, Poland;
Korea Institute of Science and Technology Information and National Research Foundation of Korea (NRF), Republic of Korea;
Ministry of Education and Scientific Research, Institute of Atomic Physics and Ministry of Research and Innovation and Institute of Atomic Physics, Romania;
Joint Institute for Nuclear Research (JINR), Ministry of Education and Science of the Russian Federation, National Research Centre Kurchatov Institute, Russian Science Foundation and Russian Foundation for Basic Research, Russia;
Ministry of Education, Science, Research and Sport of the Slovak Republic, Slovakia;
National Research Foundation of South Africa, South Africa;
Swedish Research Council (VR) and Knut \& Alice Wallenberg Foundation (KAW), Sweden;
European Organization for Nuclear Research, Switzerland;
Suranaree University of Technology (SUT), National Science and Technology Development Agency (NSDTA) and Office of the Higher Education Commission under NRU project of Thailand, Thailand;
Turkish Energy, Nuclear and Mineral Research Agency (TENMAK), Turkey;
National Academy of  Sciences of Ukraine, Ukraine;
Science and Technology Facilities Council (STFC), United Kingdom;
National Science Foundation of the United States of America (NSF) and United States Department of Energy, Office of Nuclear Physics (DOE NP), United States of America.
\end{acknowledgement}

\bibliographystyle{utphys}   
\bibliography{bibliography}

\newpage
\appendix

%
%

\section{The ALICE Collaboration}
\label{app:collab}

\begin{flushleft}

S.~Acharya$^{\rm 143}$, 
D.~Adamov\'{a}$^{\rm 98}$, 
A.~Adler$^{\rm 76}$, 
J.~Adolfsson$^{\rm 83}$, 
G.~Aglieri Rinella$^{\rm 35}$, 
M.~Agnello$^{\rm 31}$, 
N.~Agrawal$^{\rm 55}$, 
Z.~Ahammed$^{\rm 143}$, 
S.~Ahmad$^{\rm 16}$, 
S.U.~Ahn$^{\rm 78}$
% ...

\bigskip

\bigskip 

\textbf{\Large Affiliation Notes}

\bigskip 

% ...
$^{\rm V}$ Also at: Institute of Theoretical Physics, University of Wroclaw, Poland\\

\bigskip

\bigskip 

\textbf{\Large Collaboration Institutes}

\bigskip 

% ...
$^{16}$ Department of Physics, Aligarh Muslim University, Aligarh, India\\
% ...
$^{31}$ Dipartimento DISAT del Politecnico and Sezione INFN, Turin, Italy\\
% ...
$^{35}$ European Organization for Nuclear Research (CERN), Geneva, Switzerland\\
% ...
$^{55}$ INFN, Sezione di Bologna, Bologna, Italy\\
% ...
$^{76}$ Johann-Wolfgang-Goethe Universit\"{a}t Frankfurt Institut f\"{u}r Informatik, Fachbereich Informatik und Mathematik, Frankfurt, Germany\\
% ...
$^{78}$ Korea Institute of Science and Technology Information, Daejeon, Republic of Korea\\
% ...
$^{83}$ Lund University Department of Physics, Division of Particle Physics, Lund, Sweden\\
% ...
$^{98}$ Nuclear Physics Institute of the Czech Academy of Sciences, \v{R}e\v{z} u Prahy, Czech Republic\\
% ...
$^{143}$ Variable Energy Cyclotron Centre, Homi Bhabha National Institute, Kolkata, India\\
% ...

\end{flushleft} 
  
\end{document}